# Bonding Structures of $ZrH_x$ Thin Films by X-ray Spectroscopy

Martin Magnuson, Fredrik Eriksson, Lars Hultman, and Hans Högberg
*Thin Film Physics Division, Department of Physics, Chemistry and Biology (IFM), Linköping University, SE-58183 Linköping, Sweden*

## Abstract

The variation in local atomic structure and chemical bonding of $ZrH_x$ ($x$=0.15, 0.30, 1.16) magnetron sputtered thin films are investigated by Zr *K*-edge (*1s*) X-ray absorption near-edge structure and extended X-ray absorption fine structure spectroscopies. A chemical shift of the Zr *K*-edge towards higher energy with increasing hydrogen content is observed due to charge-transfer and an ionic or polar covalent bonding component between the Zr *4d* and the H *1s* states with increasing valency for Zr. We find an increase in the Zr-Zr bond distance with increasing hydrogen content from 3.160 Å in the hexagonal closest-packed metal ($\alpha$-phase) to 3.395 Å in the understoichiometric $\delta$-$ZrH_x$ film ($CaF_2$-type structure) with $x$=1.16 that largely resembles that of bulk $\delta$-$ZrH_2$. For yet lower hydrogen contents, the structures are mixed $\alpha$- and $\delta$-phases, while sufficient hydrogen loading ($x$>1) yields a pure $\delta$-phase that is understoichiometric, but thermodynamically stable. The change in the hydrogen content and strain is discussed in relation to the corresponding change of bond lengths, hybridizations, and trends in electrical resistivity.

## 1. Introduction

Metal hydrides MeH$_x$, is an interesting class of materials, where hydrogen inclusion increases the wear resistance, strength, ductility, and hardness properties of the pure metal.[1] The hydrogen atoms prevent movement of dislocations in the metal lattice,[2] which gives rise to unique properties that can be varied by changing the composition $x$ in MeH$_x$.[3] However, it is important to optimize the hydrogen content as excessive hydrogen makes the material brittle. In particular, ZrH$_x$ is useful in metallurgical applications such as hydrogen storage and as a seal between metal and ceramic materials in getter pumps in vacuum tubes[4], as hydrogenation catalyst, as neutron moderators, as well as in zircalloy[5] and in pyrotechnic initiators.[6]

The Zr-H binary phase diagram includes, as seen from an increasing hydrogen content, the stable hexagonal closest packed α–phase and a metastable ß-phase (body centered cubic) at higher temperatures.[7] Around 17 to 33 at. % H, there is evidence of a phase with trigonal crystal structure ξ-ZrH$_{0.25-0.5}$ that is fully coherent with α-Zr,[8] while a 1:1 Zr to H ratio yield the metastable γ-phase (ZrH-type structure)[9] often described by a 'face-centered tetragonal' crystal structure (8)(9)(10). Increasing the H content results in understoichiometric δ-ZrH$_x$ phase (CaF$_2$-type structure, $x$ = ~1.6-2) [10] and a body-centered tetragonal ε-phase ($x$ = 1.75–2) with ThH$_2$-type structure[11] has also been observed[10]. An understoichiometric δ-ZrH$_{2-x}$ phase has been found in both bulk[12] and thin films[13], while the tetragonal ε-phase in thin films remains to be demonstrated. Little is known experimentally about the local structure and the chemical bonds that govern the material's properties in the metal hydrides, in particular for thin films. Knowledge on how for instance impurities (oxygen) and phase mixtures affect the chemical bonding structure is beneficial to explore future applications of zirconium hydride thin films. Here, we suggest to study zirconium hydrides as an electrical contact thin film material given the higher plasticity of this electrically conducive ceramic compared to borides, carbides, and nitrides.

Previous studies of the electronic structures of ZrH$_x$ include density functional theory calculations of the δ- and ε-phases[14]. Although the density of states indicates Zr *4d* - H *1s* hybridization[15,16], experimental electronic structure studies of the Zr-H bonds are largely lacking. So far, the only electronic structure experiments were made using X-ray photoelectron spectroscopy (XPS)[17,18] on ZrH$_x$ valence band of bulk materials. A core-level study on the Zr *3d* level has also been made on thin films[13]. These spectroscopic measurements indicated charge-transfer from Zr to H that may affect the electrical properties as observed by a rather large change of the electronic states close to the Fermi level compared to pure Zr metal[12].

In this work, we investigate the local atomic structure and the trend in the chemical bonding in ZrH$_x$ thin films produced by reactive magnetron sputtering (rDCMS) as a function of average hydrogen content ($x$=0.15, 0.30, and 1.16) as well as bulk α-Zr with X-ray absorption near edge structure (XANES) and extended X-ray absorption fine structure (EXAFS) spectroscopy using synchrotron radiation and supported with X-ray diffraction (XRD) using a lab-source. Figure 1 shows the unit cells of the α-Zr and δ-ZrH$_2$ phases, as a preparation for the structure modeling, where in the δ-structure, the hydrogen atoms occupy the tetrahedral interstices in the Zr lattice. Both near-edge XANES and EXAFS with an extended energy regime contain information about the local geometry near the excited atoms and the methods complement each other. Although the scattering cross section for hydrogen is relatively low, these spectroscopies are ideal tools for characterizing the local short-range atomic coordination symmetry[19] and chemical bonding in both crystalline and amorphous materials[20,21]. XANES provides quantitative information about the average oxidation state of the absorbing element, the local coordination environment and the unoccupied electronic structure while EXAFS provides quantitative information about average bond length, the coordination numbers and type of neighbors as well as the mean-square disorder of the nearest neighboring atoms[22,23]. The XANES and EXAFS investigations are thus complemented with XRD, to assess both the short and the long-range order structural properties of the films.

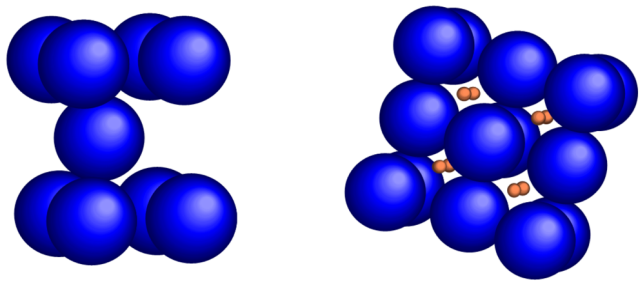

**Figure 1:** (Color online) Unit cells of a) α-Zr (space group 194) with Zr-atoms (blue spheres) in (0,0,0) and (⅓,⅔, ½) and b) stoichiometric δ-ZrH$_2$ (space group 225) with Zr-atoms (blue spheres) in (0,0,0), (½,½,0), (½,0,½), and (0,½,½) and with H (orange spheres) in all tetrahedral sites (¾,¼,¼), (¼,¾,¼), (¼,¼,¾), (¾,¾,¾), (¼,¼,¼), (¾,¾,¼), (¼,¾,¾), and (¾,¼,¼).

## 2. Experimental Details
### 2.1 Thin film deposition and characterization

The investigated Zr-H films were deposited on Si(100) substrates by rDCMS. Three ZrH$_x$ films with average compositions $x$=0.15, 0.30, and 1.16 were produced using a commercial industrial high vacuum coating system, CemeCon AG, Würselen, Germany. The growth was carried out by sputtering from a zirconium target with a purity higher than 99.9% (Hf 0.05%) using a sputtering power of 5000 W in an Ar (99.9997%)/H$_2$ (99.9996%) plasma with a fixed Ar partial pressure of 0.42 Pa and a substrate bias of -80 V. The rDCMS films were deposited for 120 s with 5, 10, and 20% H$_2$ in the plasma yielding films with thicknesses in the range ~800 to ~840 nm. For further details on the deposition conditions as well as mechanical and electrical properties of the films, the reader is referred to Ref.[13]. The investigated α-Zr sample was a commercial zirconium target with a purity of 99.9% from Kurt J. Lesker Company, Clairton, PA, USA.

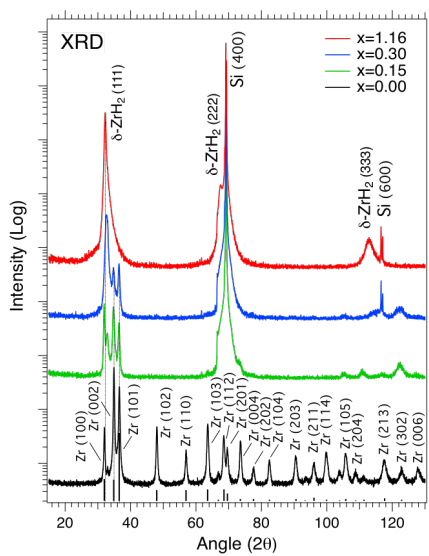

**Figure 2:** (Color online) XRD pattern of ZrH$_x$ ($x$=0.15, 0.30, and 1.16) in comparison to the diffraction pattern recorded from the α-Zr target ($x$=0). The vertical bars at the bottom represent literature values of α-Zr.[41]

Previously, time-of-flight energy elastic recoil detection analysis (ToF-ERDA) was applied to determine the hydrogen contents[13] in the films as well as their level of C, N, O, and Ar, using 36 MeV, $^{127}$I$^{8+}$ ions with a beam incidence angle of 67.5° to the sample surface normal and a recoil angle of 45°.[24] The structural properties of the films were assessed by X-ray diffraction (XRD) performing

θ/2θ scans in a Philips PW 1820 Bragg-Brentano diffractometer using Cu *Kα* radiation at settings of 40 kV and 40 mA. XRD pole figure measurements were performed using a Philips X'Pert MRD diffractometer in a parallel beam configuration with a point-focused copper anode source (Cu *Kα*, λ=1.54 Å), operating at 45 kV and 40 mA. The primary beam was conditioned using 2x2 mm$^2$ crossed-slits and in the secondary beam path a 0.27° parallel plate collimator was used together with a flat graphite crystal monochromator. A proportional detector was used for the data acquisition. The detector position was fixed at specific diffraction angles, corresponding to diffraction from δ-ZrH$_2$ {111}, {220}, and {420} family of planes, respectively. The pole figure measurements were performed in 5°-steps with azimuthal rotation 0° ≤ φ ≤ 360° and tilting 0° ≤ ψ ≤ 85° to determine the orientation distribution of the crystals. The density of the film with *x*=1.16 was determined by X-ray reflectivity (13), and found to be metallic and dense (5.91g/cm$^3$) close to the reference value for δ-ZrH$_2$ of (5.667g/cm$^3$).[25]

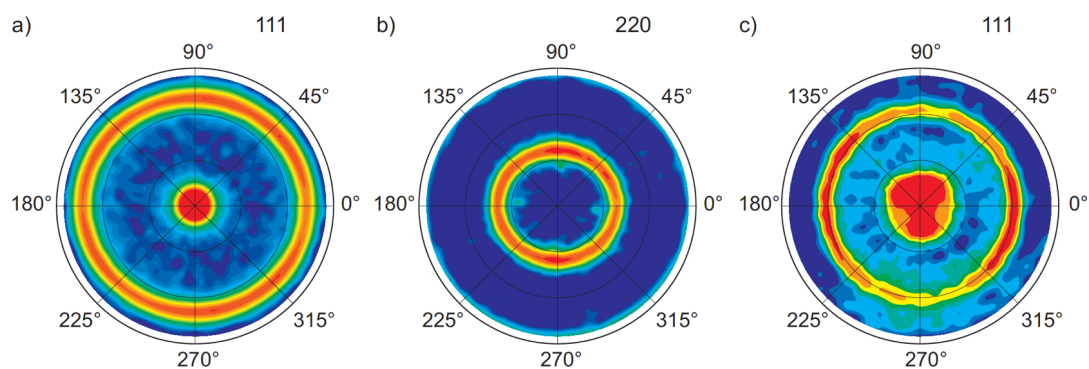

**Figure 3**: XRD pole figure measurements of the a) 111 and b) 220 reflections of ZrH$_x$ with *x*=1.16, and c) 111 reflection of ZrH$_x$ with *x*=0.15.

**2.2 XANES and EXAFS measurements**

The XANES and EXAFS spectra were measured at the undulator beamline I811 on the MAX II ring of the MAX IV Laboratory, Lund University, Sweden.[26] The beamline monochromator consists of double crystal pairs that can be altered between the Si(111) and Si(311) directions where the second crystal is bendable for horizontal focusing. The energy resolution at the Zr *1s* edge of the beamline monochromator was 0.5 eV. The X-ray absorption spectra were recorded in fluorescence mode by detecting the emitted photons using a passivated implanted planar silicon (PIPS) detector from 150 eV below to 1200 eV above the Zr *1s* absorption edge energy with 0.5 eV energy steps. To minimize self-absorption effects in the sample and Bragg scattering from the substrate, the incidence angle on the sample was normal to the sample surface and varied in 0.25° steps in a maximum range of ±3° using a stepper motor.

The Zr-H and Zr-Zr scattering paths of the first coordination shell obtained from the effective scattering amplitudes (*FEFF*)[27,28] were included in the EXAFS fitting procedure with the α-Zr and the δ-ZrH$_2$ structures as approximate model systems using the *VIPER* software package[29]. VIPER was also used to correct for the small self-absorption effects in the XAS spectra in the normal incident – grazing exit geometry for fluorescence yield.

The implemented EXAFS formula assumes random orientation or cubic symmetry while oriented samples would affect the coordination numbers depending on the polarization. The threshold energy $E_o$, is defined through the point of maximum of the first derivative at the absorption edge. In the fitting procedure, $E_o$ is used as an adjustable parameter that partly compensate for errors in the phase shifts. The threshold reference energy $E_0$ was set to the maximum of the first derivative of the pre-peak of each Zr $1s$ X-ray absorption spectrum to 17992.67 eV (α-Zr, $x=0$), 17995.96 eV ($x=0.15$), 17996.75 eV ($x=0.30$), and 18000.34 eV ($x=1.16$), respectively.

The $k^2$-weighted $\chi$ EXAFS oscillations were extracted from the raw absorption data after removing known monochromator-induced glitches and peaks originating from substrate diffraction, subsequent atomic background subtraction, and averaging of 15 X-ray absorption spectra. The bond distances ($R$), number of neighbors ($N$), Debye-Waller factors ($\sigma^2$, representing the amount of disorder) and the reduced $\chi_r^2$ as the squared area of the residual, were determined by fitting the back-Fourier-transform signal between $k=3.0-12.5$ Å$^{-1}$ originally obtained from the forward Fourier-transform within $R=1.0-3.8$ Å of the first coordination shell using a Hanning window function [27,28] and a global electron reduction factor of $S_0^2=0.8$. The thermal disorder in EXAFS is reflected in the Debye-Waller factor [30,31] that is proportional to the root-mean-square average of the difference of atomic displacements along the equilibrium bond direction.

## 3. Results and Discussion

Figure 2 shows XRD data of ZrH$_x$ films ($x=0.15$, 0.30, and 1.16) in comparison to the assigned peaks of a more randomly oriented α-Zr target ($x=0$). In the supplementary information, we show the diffraction pattern of a pure Zr reference film to explain the orientation of the α-Zr phase in the $x=0.15$ film and in the $x=0.30$ film. As observed, when the hydrogen content is gradually increasing from zero to 0.15 and then further to 0.30, the peak intensities of the most intense polycrystalline Zr peaks (100, 002, and 101) are reduced. In addition, a peak located between the 100 and 002 peaks from Zr gain intensity and overcomes that of the 002 and the 101 Zr peaks at $x=0.30$. For $x=1.16$, the Zr peaks are completely replaced by a peak located at 2θ=32.25° that is identified as the 111 peak from δ-ZrH$_2$ [25] and shows the formation of a crystalline hydride. This is an interesting observation seen from the homogeneity range determined for bulk δ-ZrH$_x$ with $x = $ ~1.6 to 2 and suggests that thin film δ-ZrH$_2$ is stable at lower hydrogen content compared to bulk material. Furthermore, the diffractogram at the top displays two additional peaks from δ-ZrH$_2$: the 222 at 2θ≈68° visible as a shoulder on the left side of the substrate 400 peak of high intensity and the 333 peak 2θ≈113° to the left of the 600-substrate peak. The 111, 222, and 333 peaks suggest an oriented film. A fiber-textured growth for this film was determined by pole figure measurements, see Figure 3a) and b). As can be seen, the monitored {111} pole in Figure 3a) displays a point of high intensity centered in the middle of the figure where ψ≈0° as well as a ring of equally high intensity at ψ≈70°. This is expected from a 111-oriented film with a cubic crystal structure that will show high intensities at the ψ-angles 0° and 70.5° [32]. The investigated {220} pole in Figure 3b supports a 111 fiber-textured growth of the film seen from a ring of uniform intensity at ψ≈35° close to the theoretical angle of 35.3° [32]. The expected intensity at ψ≈90° is only visible as scattered intensities in the periphery of Figure 3b, since the maximum measured tilt angle was $\psi=85°$. The investigated {420} pole with rings of high intensities at $\psi\approx40°$ and $\psi\approx75°$ (not shown) finally concludes that the film is 111 oriented and fiber textured. Pole figure measurements of the film with $x=0.30$ yield similar results for the {111} and {220} as previously described for the film with $x=1.16$, but with much lower intensities and where the {420} pole only display areas of scattered intensities close to the expected $\psi$-angles 39.2° and 75° [32]. This can be explained by the fact that $\delta$-ZrH$_2$ is a minority phase at this composition. As for the previous films the {111} pole figure of the film with $x=0.15$ displays intensity in the middle of the pole figure, but with a ring of uniform intensity at $\psi \approx 61°$, see Fig. 3c. The ring at $\psi \approx 61°$ probably originates from the metallic Zr contribution in this film as the angle between the 002 and 101 planes in metallic Zr is 61.5°. The diffraction pattern from the film with $x=0.15$ in Figure 2 supports this result by demonstrating clear peaks that can be assigned to Zr 002 and 101.

From the position of the 111 peak in the $\theta/2\theta$ diffractogram recorded from the ZrH$_{1.16}$ film, it was possible to determine the *a*-axis to 4.808 Å. This is slightly larger than the literature value for bulk $\delta$-ZrH$_2$ with 4.781 Å [25], but somewhat smaller than the calculated values by Chihi *et al.* [14] with $a=4.87$ Å and Zhang *et al.* [15] $a=4.823$ Å for stoichiometric ZrH$_2$. A possible explanation for the deviation from bulk is that the film is in a state of stress due to the intense flux of sputtered material applied during growth seen from a deposition rate of ~70 Å/s [13]. This high growth kinetics with energetic ions and atoms normally cause defects in thin films with excess volume (atomic peening) and surface diffusion into cusps causing compressive stress. Thus, a potential low-angle shift of the XRD peaks is caused by a combination of both changes in stoichiometry (increased hydrogen content) and macro strain in the form of compressive stress. In addition, micro strain in the form of vacancies and dislocations give rise to a peak broadening.

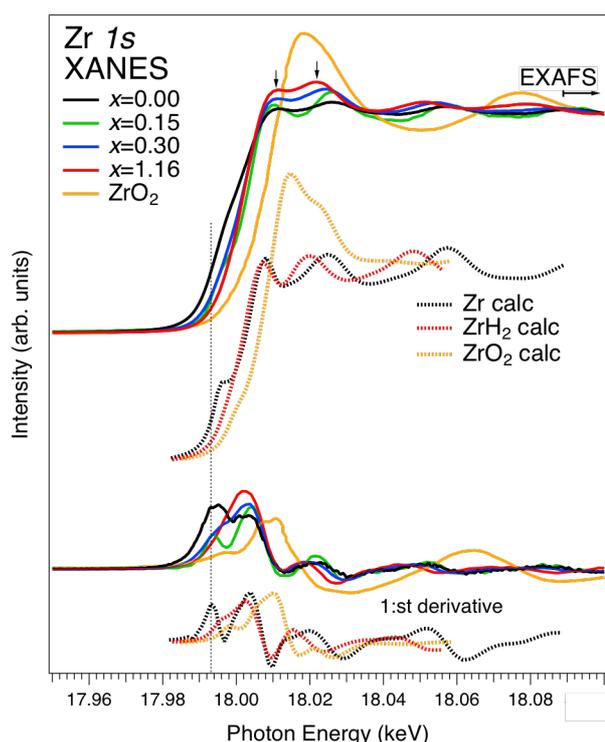

**Figure 4:** (Color online) Zr *1s* XANES spectra of ZrH$_x$ ($x=0.15$, 0.30, and 1.16), Zr metal ($x=0$) and[45] ZrO$_2$ in comparison to calculated spectra (dashed curves) of ZrH$_2$ (cubic CaF$_2$-type structure, space group=225, $a=4.781$ Å), Zr metal (hexagonal, space group=194: $a=b=3.232$ Å, $c=5.147$ Å) and ZrO$_2$ (space group 14, $a=5.149$ Å). The spectra were normalized below and above (18100 eV) the absorption edge. At the bottom, the first derivatives of the measured and calculated spectra are shown.

For comparison, a temperature-dependent *in-situ* XRD study [33] using synchrotron radiation at 87.7 keV on hydrogen loaded Zr powder, showed that $\alpha$-Zr rapidly transformed into $\delta$-ZrH$_x$ and cubic $\beta$-Zr at 313 degrees C. At higher temperature and longer time, the cubic $\beta$-Zr phase disappeared and also transformed into δ-ZrHx. Contrary to the hexagonal α-Zr phase (*a*=3.24 Å), β-Zr is cubic and has a larger lattice constant, *a*=3.76 Å. However, from XRD, we find no evidence of the high-temperature β-Zr phase in our films that were deposited without external heating.

Figure 4 (top) shows Zr *K* XANES spectra from different average compositions of ZrH$_x$ (*x*=0, 0.15, 0.30, and 1.16) in comparison to calculated spectra of hexagonal α-Zr, cubic δ-ZrH$_2$ and ZrO$_2$. The main absorption peak of α-Zr is due to pure Zr *1s → 4p* dipole transitions, while the pre-edge shoulder (pre-peak) is due to transitions into hybridized *p-d* states in elemental Zr, consistent with previous results [34,35]. As observed by the first derivative in Figure 4, the intensity of the pre-peak is most significant for pure Zr and for *x*=0.15, which is a signature of tetrahedral distortion of the coordination symmetry around the absorbing Zr atoms in the hexagonal α-Zr structure that allows *p-d* mixing into the Zr *1s → 4p* dipole transitions [36]. For ZrH$_x$, a potential pre-peak would be due to a transition of a Zr *1s* electron into mixed Zr *4d* - H *2s* states [37]. However, for *x*=0.30, the pre-peak is weaker due to smaller Zr *p-d* mixing as a result of higher octahedral symmetry for the environment around the Zr atoms in the δ-ZrH$_x$ structure. For *x*=1.16, the pre-peak has vanished and indicates that the δ-ZrH$_x$ structure is cubic. The spectral shapes and the energy shifts of calculated spectra using the FEFF code [19,27,28] in Figure 4 are generally in agreement with the experimental results. Note that the XANES spectra of α-Zr and ZrH$_x$ both have a double-structured main peak at 18.00-18.02 eV with a splitting of 11-15 eV, as indicated by the arrows. Note that the amplitude of this double peak is sharpest for *x*=0.15, specific for bulk metal, while smoother peak oscillations are observed for the other samples, specific for thin films and nanoparticles [38]. Contrary to Zr metal, the main peak of the calculated ZrO$_2$ spectrum has a different and more intense single-peak shape that occurs at higher energy and does not appear in the experimental data. This shows that our thin film samples are of high purity, which is supported from ToF-ERDA measurements [13].

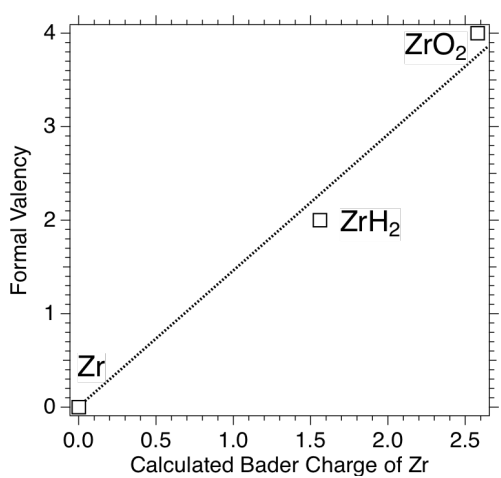

**Figure 5:** (Color online) Formal valency of elemental Zr, ZrH$_2$ and ZrO$_2$ as a function of calculated Bader charges.

The energy positions and the shapes of the main peaks and the pre-edge XANES shoulders depend on the chemical state of the absorbing atom. The bottom part of Figure 4 shows the first derivative of the absorption spectra. A general trend of the chemical peak shift is observed for the first derivative of the pre-peaks as a function of hydrogen content. The intensity of the pre-peak is gradually shifted towards higher energy and is weakened with increasing hydrogen content. The dashed lines show the corresponding calculated spectra for pure Zr, ZrH$_2$ and ZrO$_2$. The main features and trends observed in the experiment are captured in the calculated spectra of hexagonal α-Zr [36] and cubic δ-ZrH$_2$. In particular, there is a significant high-energy shift of the absorption edge for ZrO$_2$ in comparison to pure α-Zr.

Figure 5 shows the formal valency of elemental Zr, $ZrH_2$ and $ZrO_2$ as a function of calculated Bader charges. The plot highlights the differences between the chemical states of Zr in the samples as the relative energy shifts of the pre-peaks reflect the amount of *p-d* hybridization [36] and thus amount of α-Zr in the samples. Generally, the oxidation states of the absorbing Zr atoms are higher in $ZrH_x$ compared to α-Zr as the pre-peak in the absorption spectra shift to higher energies with increasing hydrogen content. Based on the energy shifts, we estimate that the Zr oxidation state increases from 0 (*x*=0), 0.22 (*x*=0.15), 0.43 (*x*=0.30) to 1.68 (*x*=1.16).

To reveal the physical origin of the polar-ionic-covalent bonding, we calculated the valence charge density distribution by Bader charges [38] using the Vienna *ab initio* simulation package (VASP) [39] of $ZrH_2$ for Zr and H. The calculated Bader charges for $ZrH_2$ and $ZrO_2$ in Fig. 5 show a linear behavior in comparison to the formal valency as approximately expected for the edge energy. In comparison to the pure Zr and H elements, the Bader charge on the Zr atoms in $ZrH_2$ increases significantly by +1.56e, *i.e.*, significant charge is withdrawn from the $4d^2 5s^2$ valence orbitals of zirconium. The charge is transferred to the lowest unoccupied orbital of the hydrogen atoms ($1s^1 \rightarrow 1s^{1.78}$), where the Bader charge decreases by -0.78e on each atom. For understoichiometric *x* < 2 in $ZrH_x$, shrinking or distortion of the cubic structure would cause a slight change in the charge transfer. For $ZrO_2$, the Bader charge of the Zr atoms further increases by +2.58e transferred to the O atoms where the Bader charge decreases by -1.29e on each atom.

Figure 6 shows experimental EXAFS oscillations in terms of structure factor as a function of the wave vector $\chi(k)$ that were $k^2$-weighted to highlight the higher *k*-region, where $k = \sqrt{2m/h^*(E-E_o)}$ is the wave vector of the excited electron. The oscillations were obtained after absorption edge determination, $E_o$ energy calibration by the first derivative, background subtraction and normalization to a spline function in *VIPER*. [29] The frequency of the oscillations and intensity of the EXAFS signal are directly related to the bond length (*R*) and the number of nearest neighbors (*N*), respectively. A higher frequency implies larger *R*, while larger amplitude of the oscillations implies increased *N*. For α-Zr and the $ZrH_x$ films, the main sharp oscillations occur in the 3-12.5 Å$^{-1}$ *k*-space region. The horizontal arrow at the top of Fig. 6 indicates this *k*-window. Note the differences in the positions and the envelopes of the oscillations. As shown, the first feature at 2.1 Å$^{-1}$ remains at the same *k*-value for *x*=0 and *x*=0.15 and only minor differences can be observed in these oscillations. For *x*=0.30 and 1.16, this feature has shifted to smaller *k*-values (1.8 Å$^{-1}$ and 1.4 Å$^{-1}$, respectively). This shift is due to a larger bond-length and the larger amplitude indicates an influence from backscattering from the light H-atoms. The same kind of differences can be observed for the peak feature at 4.1 Å$^{-1}$. The main peak at 7.1 Å$^{-1}$ due to Zr-Zr scattering becomes significantly shifted and out of phase only for *x*=1.16 in comparison to *x*=0. For *k*-values above 12.5 Å$^{-1}$, the oscillations are damped out in increasing noise. The details of the fitting procedure are described in section 2.2.

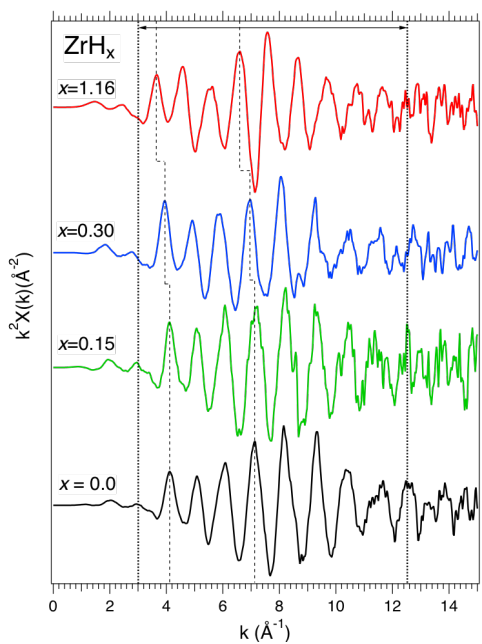

**Figure 6:** (Color online) EXAFS structure factor data S(Q) for $ZrH_x$ films for different concentrations of H in comparison to α-Zr. The fit is compared to the back-Fourier-transform of the first coordination shell within *R*=2-3.5 Å and *k*=3-12.5 Å$^{-1}$.

Figure 7 shows the magnitude of the radial distribution functions (RDF), obtained by Fourier transformation of the $k^2$-weighted $\chi(k)$ in Figure 6 by the standard EXAFS procedure.[19] The radii-values on the abscissa in Figure 7 are raw data and are not phase shifted as opposed to the refined values in Table I that shows the results of the EXAFS fitting using the scattering paths of α-Zr and δ-ZrH$_2$ as model systems. For α-Zr we applied the first two single scattering paths. For δ-ZrH$_2$, we applied the first three single scattering paths. Generally, the radial Zr-Zr distances of the magnitude of the Fourier transform of the EXAFS data for α-Zr (3.165 Å and 3.247 Å) are in good agreement with our calculated Zr-Zr bond distances (half diagonal 3.179 Å and cell edge 3.232 Å). For δ-ZrH$_x$, the Zr-H bond distances are slightly different (2.103 Å and 3.950 Å) compared to our calculated values (2.070 Å and 3.964 Å). For α-Zr, the main peak in Fig. 7 is dominated by the two Zr-Zr paths at 3.165 Å and 3.247 Å, both with six nearest neighbors in the first coordination shell (Table I). The first path (3.165 Å) corresponds to half the space diagonal in the hexagonal unit cell (see Figure 1) while the second path (3.247 Å) corresponds to the cell edge lattice parameter *a* in the hexagonal α-phase. The short-range order value of the second scattering path of α-Zr (3.247) is close to the EXAFS value reported by Edwards *et al.* (3.20 Å)[40] while our XRD yields *a*=3.233 with the literature value of 3.232 Å.[41] The fact that the second scattering path (3.247 Å) is somewhat longer than the lattice constant obtained by XRD is mainly related to the difference in probe length, as XRD records long-range order whereas EXAFS maps the short-range order in terms of mainly the 1:st coordination shell. The main difference between the scattering paths is reflected in the slightly smaller Debye-Waller factor for the first Zr-Zr path compared to the second cell edge scattering path with the same coordination number.[6] However, contrary to the cubic and isotropic δ-ZrH$_x$ structure in the films, the structure of α-Zr is hexagonal i.e., giving rise to anisotropy in the electronic structure. As a result, the D-W factor of the α-Zr contribution should be polarization-dependent. Above 3.5 Å, smaller peaks are observed due to the second and higher coordination shells.[21]

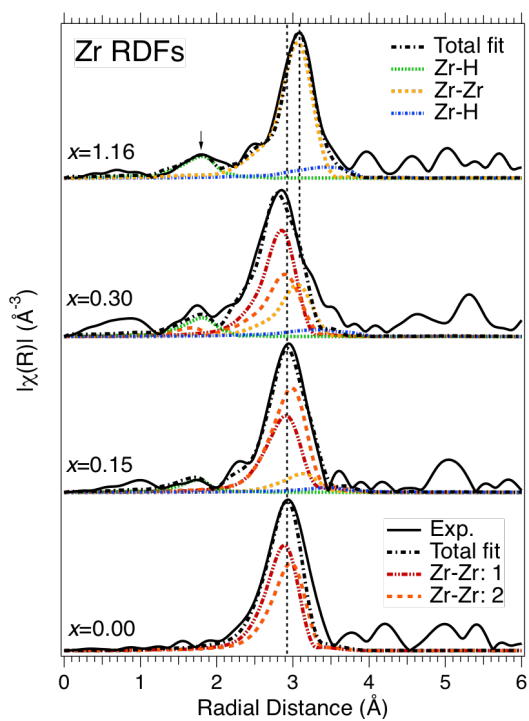

**Figure 7:** (Color online) Reduced (radial) pair distribution functions G(r) of ZrH$_x$ in comparison to pure Zr metal extracted from the Fourier transform of the EXAFS S(Q) data at the Zr *1s* absorption edge. The data is raw data, not phase shifted.

For $x=0.15$, (the ZrH$_{0.15}$ film in Figure 7), we observe a small positive shift to slightly larger distance (+0.02 Å) corresponding to an expansion of the lattice. However, for the ZrH$_{0.3}$ film there is a lattice shift to a smaller atomic distance (-0.1 Å). As observed in the EXAFS fitting, the scattering path of the first Zr-Zr shell in the α-Zr contribution has a larger coordination number and smaller Debye-Waller factor than the second shell. This gives rise to a change in the spectra weight towards lower $R$-value in Fig. 7. Furthermore, from XPS and ToF-ERDA we note that this sample ($x=0.30$) has somewhat higher oxygen content than the other samples (see fig. 4c in ref.[13]) that might lead to a more columnar microstructure and strain. As the α-Zr grain size become smaller, the interface contribution increases[42] and the relative scattering intensity from the half-diagonal coordination increase their spectral weight. The reason for strain might be an effect of changes in the stoichiometry, given the high flux of sputtered material applied during synthesis. Note that in XRD, strain due to e.g., vacancies or a more fine-grained microstructure cannot be distinguished from an eventual change in stoichiometry. For the ZrH$_{1.16}$ film, the Zr-Zr bond is further increased by +0.15 Å. For $x=1.16$, the δ-ZrH$_2$ model structure contains three different single scattering paths (listed in Table I), where the Zr-H bond length is clearly visible at 2.103 Å while the Zr-Zr path is 3.395 Å. This can be compared to our XRD results with Zr-H=2.08 Å and Zr-Zr=3.40 Å of the same sample and measured distances of Zr-H = 2.07 Å and Zr-Zr = 3.38 Å from bulk δ-ZrH$_2$.[25] The calculated bond lengths of stoichiometric ZrH$_2$ by Chihi et al. (14) are Zr-H=2.11Å and Zr-Zr=3.45 Å. Again, we suggest the differences between thin film and bulk material to be due to compressive macro strain in the former material due to the applied growth conditions.[13] We further note that for understoichiometric cubic ZrH$_{1.16}$, the refined coordination number of the nearest H neighbors for Zr (6.39) and Zr-Zr neighbors (10.10) are lower than the nominal values (N=8 and N=12) for the δ-ZrH$_2$ structure as they represent weighted averages in the structure. Using EXAFS, the ZrH$_x$ films with average compositions $x=0.15$ and $x=0.30$ (determined by ToF-ERDA) were fitted by a model structure containing a superposition of α-Zr and δ-ZrH$_2$ phases. For an estimation of the weights of the two phases, fitting constraints (N$_1$=N$_2$, N$_3$=N$_5$/3 and N$_4$=N$_5$/2) were applied with the two independent variables N$_2$ and N$_5$. The ratio N$_5$/N$_2$/4 results in approximate ratios between the phases; 28% ZrH$_x$ for $x=0.15$ and 73% for $x=0.30$. However, the error bars in the fit for the low-H samples $x=0.15$ and $x=0.30$ are higher than for the single-phase samples as they contain a superposition of two phases that leads to more parameters. Statistically, the reduced chi-squared values (chi-squared normalized by the degrees of freedom ν) are comparable (1.5-1.7).

The combined XANES and EXAFS studies at the Zr $K$-edge show several interesting effects. For the XANES spectra, the chemical shift towards higher energies in comparison to α-Zr metal is due to changes in the oxidation state that depends on the formation of Zr-H bonding and structure change at sufficient hydrogen loading. The chemical shift of the Zr $K$-edge towards higher energies confirms a significant charge-transfer from Zr towards H with increasing hydrogen content.

The trend in the charge-transfer and chemical shift with increasing hydrogen content can be correlated with increasing resistivity values. The resistivity of pure bulk α-Zr at room temperature[12] is 42.6 μΩcm while for the ZrH$_x$ films, the resistivity increases from 69 μΩcm at $x$=0.15, to 87 μΩcm for $x$=0.30 and to 116 μΩcm at $x$=1.16.[13] The increasing resistivity is due to the fact that the Zr-Zr lattice expands with the intervening non-metal hydrogen that also withdraws significant charge from the dominating Zr *4d*-orbitals close to the Fermi level.[14,15,16] The same trend in resistivity has been observed for understoichiometric bulk ZrH$_x$ materials (69.1 μΩcm for δ-ZrH$_x$ with $x$=1.54),[12] but in general, the resistivity values are higher in thin films due to factors most likely due to compressive strain, defects or a fine-grained microstructure.[13]

The EXAFS results show an expansion of the Zr-Zr bond length with hydrogen loading. From the recorded ψ-angles from the ZrH$_{1.16}$ film in the pole figure measurements, we can exclude the possibility of textured γ and ε-phases in this film. In EXAFS, the short-range coordination in the γ-phase exhibits similarities to a mixed δ-γ phase, while excessive hydrogen assembles a pure δ-phase. As observed by XRD (Figure 2), the structures resemble a superposition of the α− and δ-phases, while sufficient hydrogen loading ($x$≥1) resembles a phase-pure understoichiometric δ-phase. In fact, the δ-ZrH$_x$ phase can more easily host vacancies than other phases, and a superposition of Zr peaks with reduced intensity is therefore likely in a two-phase system.

A decrease in the Zr-Zr bond length in the understoichiometric ZrH$_x$ phase could be caused by the vacancies that separate the Zr atoms in the structure giving rise to a smaller lattice parameter, *a*. This effect can be compared to the small pressure-dependent tensile strain effect on bulk ZrH$_2$, where the Zr-Zr bond length decreases by 0.1 Å when increasing the pressure from atmospheric pressure to 13 GPa.[43] The strain caused by the vacancies in understoichiometric ZrH$_2$, distorts the bonding symmetry of the hydrogen atoms that occupy the tetrahedral sites in the cubic ZrH$_2$ lattice. For our sputtered δ-ZrH$_x$ films, we find no evidence of lattice distortion. However, for larger hydrogen contents, Jahn-Teller distortion at the tetrahedral sites favorably affects the materials properties in comparison to a tentative octahedral coordination that for hydrogen is inhibited according to the Hägg rules[44] on size and coordination number. For an octahedral coordination, the ratio between the atom in the interstitial should be ≤0.41 that of the host ion when they are arranged in the structure. For H in Zr, this ratio is 0.16 using atomic radii of 25 and 155 pm, respectively. This implies that ZrH$_x$ is tetrahedrally coordinated.

It is interesting to note that the cubic structure of our δ-ZrH$_x$ thin film is stable for lower hydrogen content compared to the corresponding bulk material. This is due to the fact that the growth of thin films for rDCMS is conducted far from the thermodynamic equilibrium. This favors growth of metastable phases outside the homogeneity range of bulk materials. The bond length obtained from EXAFS and XRD are consistent with calculated values.[14] This shows that thin film materials are applicable for analysis by XANES, EXAFS and XRD as they are comparable to bulk values. These observations imply that desired properties in ZrH$_x$ for electrical contact materials and protective coatings such as high conductivity can be controlled by the compositions of the films. In this way, other materials properties such as hardness, ductility, and wear resistance can also be improved.

## 5. Conclusions

From XANES, EXAFS and XRD, we have shown that it is possible to obtain detailed information about the local structure in reactively sputtered hydride thin films. To explore the largely unknown chemical bonding in zirconium hydrides, we study the crystal and electronic structures of $ZrH_x$ films with $x$=0.15, 0.30, and 1.16 as a potential electrical contact material with improved metallic properties in comparison to e.g., ZrC and ZrN. For low hydrogen content ($x$=0.15-0.30), the structures assemble a superposition of α– and δ-phases, but sufficient hydrogen loading ($x$=1.16) resembles a thermodynamically stable understoichiometric δ-phase. Interestingly, the thin film δ-$ZrH_x$ material is more stable at lower hydrogen content than the corresponding bulk material. The 111 fiber-textured films with hydrogen content outside the homogeneity range for bulk material structurally resembles that of a dihydride-like bulk material and shows good electrical conductivity with a resistivity of 116 μΩcm. The Zr-H and Zr-Zr bond lengths determined from XRD (2.08 Å and 3.40 Å) and EXAFS (2.103 Å and 3.395 Å) are in good agreement and we find no evidence of the metastable γ-phase from XRD or EXAFS. The Zr-H bonds are polar-covalent with a large fraction of ionic character with 1.56e transferred from the Zr to the H atoms. The trend in the charge-transfer and chemical shift with increasing hydrogen content is correlated with increasing resistivity values due to the lattice expansion and charge withdrawal from the Zr *4d*-orbitals at the Fermi level.

## Supporting Information

X-ray diffraction of a textured α-Zr zirconium reference film deposited on Si(100), using identical process conditions as for the sputtered zirconium hydride films with no hydrogen in the plasma. In order to support the orientation of the α-Zr phase determined in the $ZrH_{0.15}$ and $ZrH_{0.30}$ films in the diffraction patterns in Fig. 2.

## 7. Acknowledgements

We would like to thank the staff at MAX-IV Laboratory for experimental support. This work was supported by the Swedish Research Council (VR) Linnaeus Grant LiLi-NFM, the FUNCASE project supported by the Swedish Strategic Research Foundation (SSF). MM acknowledges financial support from the Swedish Energy Research (no. 43606-1), *the Swedish Foundation for Strategic Research (SSF)* (no. RMA11-0029) *through the synergy grant FUNCASE* and the Carl Trygger Foundation (CTS16:303, CTS14:310). HH acknowledges financial support from the Swedish Government Strategic Research Area in Materials Science on Functional Materials at Linköping University (Faculty Grant SFO-Mat-LiU No. 2009-00971).

**TABLE I:** Structural parameters for the ZrH$_{1.16}$ film in comparison to pure α-Zr obtained from fitting of calculated radial distribution functions in the first coordination shell. $N_1$ and $N_2$ are coordination numbers, $R_1$ and $R_2$ are bond length (in Å) for the first and second scattering paths for Zr-H and Zr-Zr, respectively, $\sigma_1$ and $\sigma_2$ are the corresponding Debye-Waller factors representing the amount of atomic displacement and disorder, reduced $\chi_r^2$ as the squared area of the residual. Degrees of freedom: ν=N-P, N=number of independent measurement points, P=number of fitting parameters.

| System | Shell | R(Å) | N | σ(Å$^2$) | ΔE$_0$(eV) | Statistics |
|---|---|---|---|---|---|---|
| ZrH$_{1.16}$ | Zr-H (N=8) | 2.103±0.005 | 6.39±0.01 | 0.0021±0.001 | 1.08±0.01 | $\chi_{16}^{0.95}$ = 25.73 |
| | Zr-Zr (N=12) | 3.395±0.005 | 10.10±0.01 | 0.0092±0.001 | 0.58±0.01 | N=28, P=12 |
| | Zr-H (N=24) | 3.950±0.005 | 19.03±0.01 | 0.0093±0.001 | 1.08±0.01 | ν=N-P=16 |
| ZrH$_{0.30}$ | Zr-H (N=8) | 2.103±0.005 | 2.989±0.01 | 0.0021±0.001 | 0.10±0.01 | $\chi_{24}^{0.95}$ = 35.8 |
| | Zr-Zr (N=12) | 3.375±0.005 | 2.031±0.01 | 0.0089±0.001 | 0.11±0.01 | N=28, P=4 |
| | Zr-H (N=24) | 3.971±0.005 | 17.03±0.01 | 0.0093±0.001 | 0.10±0.01 | ν=N-P=24 |
| | Zr-Zr (N=6) | 3.160±0.005 | 3.149±0.01 | 0.0069±0.001 | 0.11±0.01 | |
| | Zr-Zr (N=6) | 3.260±0.005 | 2.930±0.01 | 0.0072±0.001 | 0.11±0.01 | |
| ZrH$_{0.15}$ | Zr-H (N=8) | 2.101±0.005 | 1.010±0.01 | 0.0021±0.001 | 2.50±0.01 | $\chi_{10}^{0.95}$ = 15.51 |
| | Zr-Zr (N=12) | 3.375±0.005 | 1.700±0.01 | 0.0080±0.001 | 2.90±0.01 | N=28, P=18 |
| | Zr-H (N=24) | 3.961±0.005 | 15.02±0.01 | 0.0083±0.001 | 2.50±0.01 | ν=N-P=10 |
| | Zr-Zr (N=6) | 3.160±0.005 | 3.930±0.01 | 0.0056±0.001 | 2.90±0.01 | |
| | Zr-Zr (N=6) | 3.254±0.005 | 5.030±0.01 | 0.0059±0.001 | 2.90±0.01 | |
| α-Zr | Zr-Zr (N=6) | 3.165±0.005 | 6.00 | 0.0044±0.001 | 0.67±0.01 | $\chi_8^{0.95}$ = 14.06 |
| | Zr-Zr (N=6) | 3.247±0.005 | 6.00 | 0.0051±0.001 | 0.67±0.01 | N=12, P=4 |
| | | | | | | ν=N-P=8 |

**Figure 8:** (Color online) Unit cells of a) α-Zr (space group 194) with Zr-atoms (blue spheres) in (0,0,0) and (⅓,⅔, ½) and b) stoichiometric δ-ZrH$_2$ (space group 225) with Zr-atoms (blue spheres) in (0,0,0), (½,½,0), (½,0,½), and (0,½,½) and with H (orange spheres) in all tetrahedral sites (¾,¼,¼), (¼,¾,¼), (¼,¼,¾), (¾,¾,¾), (¼,¼,¼), (¾,¾,¼), (¼,¾,¾), and (¾,¼,¼).

**Figure 9:** (Color online) XRD pattern of ZrH$_x$ ($x$=0.15, 0.30, and 1.16) in comparison to the diffraction pattern recorded from the α-Zr target ($x$=0). The vertical bars at the bottom represent literature values of α-Zr.[41]

**Figure 10**: XRD pole figure measurements of the a) 111 and b) 220 reflections of ZrH$_x$ with $x$=1.16, and c) 111 reflection of ZrH$_x$ with $x$=0.15.

**Figure 11:** (Color online) Zr *1s* XANES spectra of ZrH$_x$ ($x$=0.15, 0.30, and 1.16), Zr metal ($x$=0) and[45] ZrO$_2$ in comparison to calculated spectra (dashed curves) of ZrH$_2$ (cubic CaF$_2$-type structure, space group=225, $a$=4.781 Å), Zr metal (hexagonal, space group=194: $a$=$b$=3.232 Å, $c$=5.147 Å) and ZrO$_2$ (space group 14, $a$=5.149 Å). The spectra were normalized below and above (18100 eV) the absorption edge. At the bottom, the first derivatives of the measured and calculated spectra are shown.

**Figure 12:** (Color online) Formal valency of elemental Zr, ZrH$_2$ and ZrO$_2$ as a function of calculated Bader charges.

**Figure 13:** (Color online) EXAFS structure factor data S(Q) for ZrH$_x$ films for different concentrations of H in comparison to α-Zr. The fit is compared to the back-Fourier-transform of the first coordination shell within $R$=2-3.5 Å and $k$=3-12.5 Å$^{-1}$.

**Figure 14:** (Color online) Reduced (radial) pair distribution functions G(r) of ZrH$_x$ in comparison to pure Zr metal extracted from the Fourier transform of the EXAFS S(Q) data at the Zr *1s* absorption edge. The data is raw data, not phase shifted.